\documentclass[5p]{elsarticle}

\usepackage{lineno,hyperref}

\journal{}

\biboptions{authoryear}

\usepackage{tikz,pgfplots}
\pgfplotsset{compat=newest} 
\usetikzlibrary{shapes,arrows,positioning}
\usepackage{amsmath}
\usepackage{siunitx}
\DeclareSIUnit\cell{cell}
\usepackage{amssymb}

\newtheorem{theorem}{Theorem}

\newtheorem{lemma}{Lemma}
\newtheorem{proposition}{Proposition}
\newproof{prf}{Proof}

\hypersetup{
  colorlinks   = true, 
  urlcolor     = cyan, 
  linkcolor    = cyan, 
  citecolor   = cyan 
}

\definecolor{mycolor1}{rgb}{0.00000,0.44700,0.74100}%
\definecolor{mycolor2}{rgb}{0.92900,0.69400,0.12500}
\definecolor{mycolor3}{rgb}{0.85000,0.32500,0.09800}%
\definecolor{grey1}{rgb}{0.5,0.5,0.5}%
\definecolor{grey2}{rgb}{0.7,0.7,0.7}%
\definecolor{grey3}{rgb}{0.9,0.9,0.9}%

\begin{document}

\begin{frontmatter}

\title{A resource dependent protein synthesis model for evaluating synthetic circuits} 

\author[ISTaddress]{Wolfgang Halter\fnref{myfootnote}}
\author[ISTaddress]{Jan Maximilian Montenbruck}
\author[ISTaddress]{Zoltan A. Tuza}
\author[ISTaddress]{Frank Allg\"ower}

\address[ISTaddress]{Institute for Systems Theory and Automatic Control, University of Stuttgart, Pfaffenwaldring 9, Stuttgart, Germany}
\fntext[myfootnote]{Corresponding author.}

%
%

\begin{abstract}
Reliable in-silico design of synthetic gene networks necessitates novel approaches to model the process of protein synthesis under the influence of limited resources. We present such a novel protein synthesis model which originates from the Ribosome Flow Model and among other things describes the movement of RNA-polymerase and Ribosomes on mRNA and DNA templates respectively. By analyzing the convergence properties of this model based upon geometric considerations we present additional insights into the dynamic mechanisms of the process of protein synthesis. Further, we exemplarily show how this model can be used to evaluate the performance of synthetic gene circuits under different loading scenarios.
\end{abstract}


\end{frontmatter}


\section{Introduction}
One of the major issues in the field of synthetic biology is the gap between the computationally predicted performance of a synthetic circuit and the performance observed in its implementation in-vitro. This problem mainly stems from the fact that assumptions made during the modeling process are oversimplifying the dynamics of the biological processes under study. Genetic regulatory networks usually have been described focusing on the direct interactions between genes and their products, neglecting the fact that there exist significant indirect couplings between all genes, including the ones not modeled. Some of these couplings originate from the usage of shared resources of the transcriptional and translational machinery. The influence of such limited pools of resources has been addressed just recently by \cite{Gyorgy2015,Weisse2015,Gorochowski2016}, where both experimental and computational approaches are being discussed. For the purpose of describing interactions of several genes and their products, the stated works mainly use Hill-kinetics to phenomenologically describe protein production depending on the concentration of certain transcription factors. The process of protein synthesis however can be described and modeled on various levels of detail and a more mechanistic approach would be beneficial in order to understand the system on a microscopic level and better evaluate the degrees of freedom for possible modifications in terms of the design of synthetic gene circuits. Particularly, considering translational control as an additional mechanism for genetic interactions may yield one possible strategy to avoid negative effects of limited pool resources. A suitable protein synthesis model should therefore allow for the implementation of different genetic control mechanisms such as transcriptional and translational control but also incorporate limitations of available resources within the cell. Therefore, the process of protein synthesis can be described as a sequence of several steps, which in turn are described on a low level of detail: transcription initiation, mRNA elongation, translation initiation and protein elongation. Post-translational modification will be neglected for simplicity. This way the resulting model satisfies the just stated requirements while remaining computationally tractable. It is also in accordance with the results of \cite{Ben-Taboude-Leon2009}, in which the authors claim that there are only two factors limiting the transcription rate: transcription initiation rate and RNA polymerase (RNAP) translocation rate. This is due to the fact that RNAP needs to proceed a certain length before the next RNAP can bind. The length of a gene then determines the dead time and transcription rate for one bound RNAP. The initiation rate on the other hand mainly depends on the strength of promoter as well as the presence of certain transcription factors. In case of translation, the physical mechanisms are assumed to be similar. Instead of RNAP, the translocation of Ribosomes and initiation of translation, which is now dependent on the strength of the Ribosome Binding Site, are the rate limiting factors. Particularly, \cite{Raveh2016} offer an approach to model the process of translation on this desired level of mechanistic detail. While they consider the flow of Ribosomes on a single mRNA template however, an extension to capturing both transcription and translation is desired and will be presented in the remainder of this work. \newline
After introducing this novel model we study its convergence properties based on geometric considerations in order to shed some light on the system theoretic properties of the model. The presented analysis therefore not only characterizes limiting sets, but further provides insights into the dependencies of the dynamics and steady states on certain parameters of the model. Subsequently, two application examples are provided which show that the new protein synthesis model based on the Ribosome Flow Model can be used to describe the basal transcriptional and translational load of a desired host organism. Therefore, the interaction between the basal activity and newly introduced synthetic circuits can be evaluated, which is shown in the second application example.

\begin{table*}[htbp]
\renewcommand{\arraystretch}{1.3}
\centering
\begin{tabular}{l l}
\hline
$R_{\text{rib,tot}} \in [0,\infty)$& total molecular amount of Ribosomes \\
$R_{\text{rib}} \in [0,\infty)$& molecular amount of free Ribosomes \\
$z_{i}  \in [0,1]$& avg. Ribosome density at mRNA location $i = 1,\ldots, m$\\
$P \in [0,\infty)$& molecular amount of protein \\
$\eta \in \mathbb{R}^+$& translation initiation rate\\
$\eta_c  \in \mathbb{R}^+$& translation elongation rate\\ 
$\delta  \in \mathbb{R}^+$& protein degradation rate\\ 
$m \in \mathbb{N}$& number of discretization points on mRNA template\\
\hline
\end{tabular} 
\caption{States and parameters of the process of translation.}\label{tab:RFMstate}
\end{table*}

\section{Protein synthesis model} \label{sec:model}
To describe the movement of Ribosomes and RNAP along a template of mRNA and DNA respectively, usually the totally asymmetric exclusion process (TASEP) is applied \citep{Shaw2003}. As this is a stochastic model of infinite dimension and therefore does not satisfy our demands for computational tractability, one usually employs a simplified deterministic version of the TASEP, obtained by a mean field approximation now also known as Ribosome Flow Model (RFM) \citep{Reuveni2011, Edri2014, Raveh2016}. So far, a combined model of transcription and translation using such probabilistic flow models does not exist. We first extend the RFM from \cite{Raveh2016} where only single mRNA templates are considered to the case where several mRNAs of the same kind are present. Subsequently, the processes of transcription and translation are coupled such that the product of transcription, mRNA, is the template for translation, therefore making the number of mRNA templates a state of our dynamical system instead of a static variable.

\subsection{Translation model}
The original RFM, introduced by \cite{Reuveni2011} was extended by \cite{Raveh2016} to also consider a finite pool of Ribosomes which partly may be bound to the mRNA template. In greater detail, translation is initiated by the Ribosome binding to the mRNA, subsequently it moves along the mRNA until it reaches the end and unbinds again. The movement is unidirectional, meaning no backward flow of Ribosomes is possible. Further, several Ribosomes may be bound to one template as long as the mRNA is long enough. In general, the speed of forward motion is not constant but dependent on the codon which is translated, or to be more precise, the available amount of tRNAs for the necessary amino acid. For the sake of simplicity however, a constant elongation speed is assumed in the remainder.\newline
The process of translation including the dynamics of the Ribosomes can be modeled as the set of differential equations
\begin{linenomath}
\begin{align}
\begin{bmatrix}
\dot{z}_1 \\
\dot{z}_2 \\
\vdots\\
\dot{z}_m 
\end{bmatrix} &= \begin{bmatrix}
\eta R_{\text{rib}} (1-z_1) - \eta_c z_1 (1-z_2) \\
\eta_c z_1 (1-z_2) - \eta_c z_2 (1-z_3)\\
\vdots\\
\eta_c z_{m-1} (1-z_m) - \eta_c z_m
\end{bmatrix} \label{eqn:singleRFM_1}\\
\dot{P} &= \eta_c z_m - \delta P,\label{eqn:singleRFM_2}
\end{align} 
\end{linenomath}
with the variable 
\begin{linenomath}
\begin{align}
R_{\text{rib}}&=R_{\text{rib,tot}} - \sum_{i=1}^m z_i \label{eqn:singleRFM_3}
\end{align}
\end{linenomath}
and all initial conditions set to zero. The model states and parameters are defined in Table \ref{tab:RFMstate}.\newline
For a specific mRNA template, the number of discretization points $m$ is determined such that each state represents the length on the mRNA lattice which is occupied by one Ribosome, therefore
\begin{linenomath}
\begin{equation}
m = \frac{L_M}{L_{\text{rib}}}
\end{equation} 
\end{linenomath}
with $L_M$ the total length of the mRNA template and $L_{\text{rib}}$ the specific length a single Ribosome occupies on the mRNA template. Typically, the elongation rate $\eta_c$, the total amount of available Ribosomes $R_{\text{rib,tot}}$ and the specific size of Ribosomes $L_{\text{rib}}$ are either constant or dependent on systemic parameters such as the availability of tRNA, temperature or growth conditions and therefore, these parameters might be determined for a specific cell type through biological experiments. The remaining parameters, namely the length of mRNA template $L_M$, the initiation rate $\eta$ as well as the protein degradation rate $\delta$ are all depending on the specific mRNA template (gene) and can thus also be considered as design parameters for the synthesis of genetic circuits.

\subsection{Multiple templates}
In the original works on the RFM with pool \citep{Raveh2016}, several mRNA templates are considered to interact with the pool of Ribosomes and each template is modeled as one individual RFM. This means, the dynamics of several templates can be observed also in cases where the initiation of translation on two identical templates is happens at different points in time. In the following, we will omit this case in order to describe identical templates with a single RFM and thus reduce the computational burden and complexity of the model. The following assumption is therefore imposed: \newline
Given several identical mRNA templates which interact with a pool of Ribosomes, the dynamic behavior of the movement of Ribosomes on the templates is identical for all templates. In other words this means that, given two mRNA templates and the dynamics of Ribosomes flowing on these templates is described by $\mathbf{z}^A(t)$ and $\mathbf{z}^B(t)$, we assume that their initial conditions satisfy
\begin{linenomath}
\begin{align}
\mathbf{z}^A(0) = \mathbf{z}^B(0).
\end{align}
\end{linenomath}
If this is the case, the model equations for a single RFM with pool (\ref{eqn:singleRFM_1})-(\ref{eqn:singleRFM_3}) can be extended to the case of a RFM with pool with $M$ identical templates by adapting the protein production rate (\ref{eqn:singleRFM_2}) and the amount of available resources (\ref{eqn:singleRFM_3}), yielding
\begin{linenomath}
\begin{align}
\dot{P} &= M \eta_c z_m - \delta P, \label{eqn:multiRFM_1} \\
R_{\text{rib}}&=R_{\text{rib,tot}} - M \sum_{i=1}^m z_i.
\end{align} 
\end{linenomath}

\begin{table*}[htbp]
\renewcommand{\arraystretch}{1.3}
\centering
\begin{tabular}{l l}
\hline
$R_{\text{rnap,tot}} \in [0,\infty)$& total molecular amount of RNAP \\
$R_{\text{rnap}} \in [0,\infty)$& molecular amount of free RNAP \\
$x_{i}  \in [0,1]$& avg. RNAP density at DNA location $i = 1,\ldots, n$\\
$G \in [0,\infty)$& Gene copy number \\
$M \in [0,\infty)$& molecular amount of mRNA \\
$\lambda \in \mathbb{R}^+$& transcription initiation rate\\
$\lambda_c  \in \mathbb{R}^+$& transcription elongation rate\\ 
$\nu  \in \mathbb{R}^+$& mRNA degradation rate\\ 
$n \in \mathbb{N}$& number of discretization points on DNA template\\
\hline
\end{tabular} 
\caption{States and parameters of the process of transcription.}\label{tab:RNAPFMstate}
\end{table*}

\subsection{Transcription model}
\begin{figure}
  \centering
  \resizebox {\columnwidth} {!} {
  \begin{tikzpicture}[scale=1,>=triangle 45]
	\node (A) at (-.7,0) [rectangle,thick,draw,anchor=east] {$\begin{aligned}\dot{\mathbf{x}} &= f_1(\mathbf{x},\lambda,\nu) \\ y_1 &= M \end{aligned}$};	
	\node (B) at (.7,0) [rectangle,thick,draw,anchor=west] {$\begin{aligned}\dot{\mathbf{z}} &= f_2(\mathbf{z},\eta,M) \\y_2 &= P \end{aligned}$};	
    \draw[<-,thick] (A.west)  --++ (180:1cm) node(L){} node[above right,xshift=3pt]{$\lambda, \nu$};
    \draw[->,thick] (A.east) node[above right,xshift=2pt]{$M$} -- (B.west) ;
    \draw[->,thick] (B.east) node[right](R){} node[above right,xshift=3pt]{$P$} --++ (0:1cm) ;
    \draw[<-,thick] (B.195)  --++ (180:.5cm) --++ (270:1cm) node[above left]{$\eta$};
    
    \draw[dashed] (-0.6,0) --++ (0,-1) -| (L.east) --++ (0,1)  -| (-0.6,0) node[pos=0.25, above]{Subsystem A} ;
    
    \draw[dashed] (0.3,0) --++ (0,-1) -| (R) --++ (0,1)  -| (0.3,0) node[pos=0.25, above]{Subsystem B} ;
  \end{tikzpicture}
  }
  \caption{Block representation of the proposed protein synthesis model.}
  \label{fig:CFM_blocks}
\end{figure}
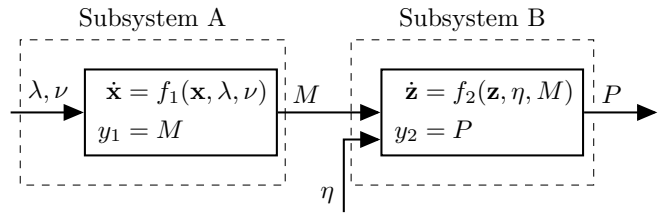

In order to arrive at a combined model of transcription and translation, it is left to define the transcriptional model and establish the connections to the translational part. \cite{Edri2014} introduced a transcription model of the RFM where the main difference lies in the fact that the RNAP is allowed to flow in both directions on the gene template and therefore different RNAP density profiles at steady state are obtained. However, as experimental evidence for this bidirectional movement is lacking, this additional mechanism will be neglected and we assume that both processes of transcription and translation can be described by the same mechanisms, only differing in the nature of the template and the pool of resources. The number of discretization points $n$ of the transcription model is obtained similarly as before, namely as the fraction of total length of the gene $L_G$ and the specific length of a RNAP $L_{\text{rnap}}$, i.e.,
\begin{linenomath}
\begin{align}
n=\frac{L_G}{L_{\text{rnap}}}.
\end{align}
\end{linenomath}
It is noted that a gene and the respective mRNA template do not need to be of the same length in this framework (and in nature as well), however, for the sake of simplicity we will assume that these templates have the same amount of codons in the remainder.\newline
For the combined model, it is therefore only necessary to extend the model states and parameters from Table \ref{tab:RFMstate} with the ones in Table \ref{tab:RNAPFMstate}. The product of the transcription model is the molecular amount of mRNA, $M$, which serves as template for the translation model and thus the template numbers are a state of our dynamical system instead of a static variable. Further, there are now two laws of mass conservation for the total amount of Ribosomes and RNAP repectively, viz.
\begin{linenomath}
\begin{align}
R_{\text{rnap}}&=R_{\text{rnap,tot}} - G \sum_{i=1}^n x_i \\
R_{\text{rib}}&=R_{\text{rib,tot}} - M \sum_{i=1}^m z_i .
\end{align}
\end{linenomath}
These equations can also be expressed as differential equations which is more compliant with the overall notation. With these states and parameters defined, the equations for the combined transcription and translation model are expressed as
\begin{linenomath}
\begin{align}
\dot{R}_{\text{rnap}}&= -G \lambda R_{\text{rnap}} (1-x_1) + G \lambda_c x_n \label{eqn:CFM_first}\\
\begin{bmatrix}
\dot{x}_1 \\
\dot{x}_2 \\
\vdots\\
\dot{x}_n 
\end{bmatrix} &= \begin{bmatrix}
\lambda R_{\text{rnap}} (1-x_1) - \lambda_c x_1 (1-x_2) \\
\lambda_c x_1 (1-x_2) - \lambda_c x_2 (1-x_3)\\
\vdots\\
\lambda_c x_{n-1} (1-x_n) - \lambda_c x_n
\end{bmatrix} \label{eqn:CFM_second}\\
\dot{M} &= G \lambda_c x_n - \nu M \label{eqn:CFM_third}\\
\dot{R}_{\text{rib}}&=-\dot{M}\sum_{i=1}^m z_i -M \eta R_{\text{rib}} (1-z_1) + M \eta_c z_m \label{eqn:CFM_fourth} \\
\begin{bmatrix}
\dot{z}_1\\
\dot{z}_2\\
\vdots \\
\dot{z}_m
\end{bmatrix} &= \begin{bmatrix}
\eta R_{\text{rib}} (1-z_1) - \eta_c z_1 (1-z_2) \\
\eta_c z_1 (1-z_2) - \eta_c z_2 (1-z_3)\\
\vdots\\
\eta_c z_{m-1} (1-z_m) - \eta_c z_m
\end{bmatrix} \label{eqn:CFM_fifth}\\
\dot{P} &= M \eta_c z_m - \delta P, \label{eqn:CFM_sixth}
\end{align}
\end{linenomath}
with initial conditions of all states equal to zero except
\begin{linenomath}
\begin{align}
R_{\text{rnap}}(t=0) &= R_{\text{rnap,tot}}\\
R_{\text{rib}}(t=0) &= R_{\text{rib,tot}}.
\end{align}
\end{linenomath}
For simplicity of notation, we will group the variables such that
\begin{linenomath}
\begin{align}
\mathbf{x} &= \begin{bmatrix}
R_{\text{rnap}} & x_1 & x_2 & \ldots & x_n & M 
\end{bmatrix}^\top \\
\mathbf{z} &= \begin{bmatrix}
R_{\text{rib}} & z_1 & z_2 & \ldots & z_m & P 
\end{bmatrix}^\top
\end{align}
\end{linenomath}
and
\begin{linenomath}
\begin{align}
\dot{\mathbf{x}} &= f_1(\mathbf{x}) \label{eqn:sysA}\\
\dot{\mathbf{z}} &= f_2(\mathbf{x},\mathbf{z}). \label{eqn:sysB}
\end{align}
\end{linenomath}
As the states $x_1,\ldots,x_n$ and $z_1,\ldots,z_m$ are probability densities, it only makes sense to consider solutions
\begin{linenomath}
\begin{align}
t &\mapsto \begin{bmatrix}
\mathbf{x}(t) \\ \mathbf{z}(t)
\end{bmatrix} 
\end{align}
\end{linenomath}
for which for all times 
\begin{linenomath}
\begin{align}
\mathbf{x}(t) & \in \Omega_x := [0,R_{\text{rnap,tot}}] \times [0,1]^n \times [0,\infty),\\
\mathbf{z}(t) & \in \Omega_z := [0,R_{\text{rib,tot}}] \times [0,1]^m \times [0,\infty).
\end{align}
\end{linenomath}
This concludes the definition of our novel protein synthesis model which not only describes the production of protein dependent on the transcription initiation rate, but also considers further physiological parameters and gives insights into the amount and dynamics of Ribosomes and RNAP bound to mRNA and DNA respectively. \newline
Further, when considering interactions between genes and their products, different control mechanisms such as transcription factor control, Riboswitches and silencing RNAs are possible. These control mechanisms manipulate the transcription initiation parameter $\lambda$, translation initiation parameter $\eta$ and the degradation rate $\nu$ of available mRNA, respectively. Therefore these variables are considered as inputs to the protein synthesis model, while the amount of protein is considered as output of the system. All remaining parameters, such as the number of gene templates $G$ or translation and transcription elongation rates are design variables and assumed to be constant over time. Figure \ref{fig:CFM_blocks} depicts a block representation of the protein synthesis model with the said inputs and outputs. Therein, $f_1$ is defined by (\ref{eqn:sysA}) and $f_2$ by (\ref{eqn:sysB}). 
\section{Equilibria and their stability properties}
With the combined model for transcription and translation at hand, it is now possible to study system theoretic properties of this model. In order to do so, one might be tempted to build on existing results on the single RFM \citep{Margaliot2012} or the RFM with pool \citep{Raveh2016}, which both are based on monotone systems theory. However, these approaches are not applicable to the combined model as the monotonicity property is not preserved. This is found by noticing that (\ref{eqn:sysA})-(\ref{eqn:sysB}) is not cooperative (cf. \cite{Hirsch1982}).
\begin{proposition}
The combined protein synthesis model \\ given by equations (\ref{eqn:sysA})-(\ref{eqn:sysB}) is neither cooperative nor competitive.
\end{proposition}
\begin{prf}
It suffices to examine two off-diagonal elements of the Jacobian of (\ref{eqn:sysA})-(\ref{eqn:sysB}) and show that they can have different signs. For instance
\begin{linenomath} \begin{align}
\frac{\partial}{\partial x_n} \dot{M} &= G \lambda_c &> 0 \\
\frac{\partial}{\partial x_n} \dot{R}_{\text{rib}} &= -G \lambda_c \sum_{i=1}^m z_i &< 0 
\end{align} \end{linenomath}
for at least one $z_i$ not being zero are a suitable choice and thus conclude the proof. \qed
\end{prf}
This further means that the flow of our system is not monotone though monotonicity is usually observed in RFM models (cf. \cite{Margaliot2012}). Rather than using monotone systems theory, we introduced a geometric approach to study the convergence properties of the RFM with pool in \cite{Halter2016} and the ideas established there will be extended in the remainder of this section. \newline
After briefly reviewing our results from \cite{Halter2016} we show that these results also hold for multiple templates and the additional output $M$ in order to investigate the stability properties of equilibria of system (\ref{eqn:sysA}) (Subsystem A of Figure \ref{fig:CFM_blocks}). Subsequently we extend this approach to the combined model.

\subsection{Flow models with multiple templates}
In \cite{Halter2016}, we showed that the equilibria of the RFM with pool, given a single mRNA template, constitute a normally hyperbolic invariant submanifold $\gamma((0,\bar{s}))$, with $\gamma$ being some curve, which is asymptotically stable. In order to prove normal hyperbolicity, it was first shown that the Jacobian matrix of the vector field under study evaluated at $\gamma(s)$ has exactly one eigenvalue equal to zero and all remaining eigenvalues strictly smaller than zero. The eigenvector which is associated with the zero eigenvalue further is lineraly dependent on $\frac{d}{ds}\gamma(s)$ or in other words lies in the tangent space of $\gamma((0,\bar{s}))$. Thus, $\gamma((0,\bar{s}))$ is a submanifold of equilibria. Additional invariant affine subspaces $S_p$ are constituted due to the mass conservation of the Ribosomes and it can be shown that these $S_p$ intersect $\gamma$ uniquely and transversely. This reveals normal hyperbolicity as the whole state space can be continuously split up into the tangent space $T_{\gamma(s)}\gamma((0,\bar{s}))$ and the stable normal space $N^s$ which is spanned by the remaining eigenvectors of the Jacobian. Asymptotic stability then follows directly. For further details on this proof, the reader is referred to \cite{Halter2016}.\newline
The differences between the system studied in \cite{Halter2016} and Subsystem A of Figure \ref{fig:CFM_blocks} are that the number of templates $G$ is now allowed to take any other value in $\mathbb{N}$ and that we further have an additional state $M$ as an output of the system. Therefore we first show that the results of \cite{Halter2016} still hold for these extensions.\newline
Similar to \cite{Halter2016} we first bring (\ref{eqn:sysA}) into the form
\begin{linenomath} \begin{align}
\dot{\mathbf{x}} = A(\mathbf{x})\mathbf{x}
\end{align} \end{linenomath}
with
\begin{linenomath} \begin{equation}
\begin{split}
& \hspace{-.4cm} A(\mathbf{x}) = \\ & \resizebox{.44 \textwidth}{!}
{$\begin{bmatrix}
-G \lambda (1-x_1) & 0 & \cdots & 0 & G \lambda_c & 0\\
\lambda (1-x_1) & -\lambda_c (1-x_2) & 0 & \cdots & 0 & 0\\
0 & \lambda_c (1-x_2) & -\lambda_c (1-x_3) &0 & \vdots & \vdots\\
\vdots & 0 & \ddots &  \ddots& 0 & 0\\
0& \cdots&  0 & \lambda_c (1-x_{n}) & -\lambda_c & 0\\
0& \cdots&  0 & 0 & G \lambda_c & -\nu\\
\end{bmatrix}
$}
\end{split}
\end{equation} \end{linenomath}
in order to find a parameterization $\gamma: s \mapsto \gamma(s)$ of the equilibria. Thus we define $\gamma$ such that
\begin{linenomath} \begin{align}
\forall \mathbf{x} \in \text{int }\Omega_x  \cap \ker A(\mathbf{x}) \quad \exists s \in (0,\bar{s})& \quad \mathbf{x} =\gamma(s) \label{eqn:gammadef}
\end{align} \end{linenomath}
and find that
\begin{linenomath} \begin{equation}
\gamma: s \mapsto \begin{bmatrix}
\gamma_{0}(s) & \ldots & \gamma_{n+1}(s) 
\end{bmatrix}^{\top}
\end{equation} \end{linenomath}
with the components $\gamma_{i}(s)$ given recursively as a series of continued fractions with
\begin{linenomath} \begin{equation}
\gamma_{i}(s) = \begin{cases}
\tfrac{\lambda_c s}{\lambda(1-\gamma_{1}(s))} & i = 0\\
 \tfrac{s}{1-\gamma_{i+1}(s)} & i = 1 \ldots (n-1)\\
 s & i=n \\
\tfrac{G\lambda_c}{\nu}s & i=n+1.
 \end{cases} \label{eqn:gamma_recu}
\end{equation} \end{linenomath}
We note that the restriction to $\mathbf{x} \in \text{int }\Omega_x$ and therefore also the upper bound $\bar{s}$ in $s \in (0,\bar{s})$ is rather technical and has sufficiently been discussed in \cite{Halter2016}.\\
With this representation of the equilibria at hand we can now study the Jacobian of $f_1$ evaluated at $\gamma(s)$, i.e.
\begin{linenomath} \begin{equation}
\begin{split}
& \hspace{-.4cm} J_{f_1}(\gamma)= \\ & \resizebox{.44 \textwidth}{!}
{$
\begin{bmatrix}
-G \lambda(1-\gamma_1) & G \lambda \gamma_0                        & 0                                & \cdots   & G \lambda_c & 0\\
\lambda(1-\gamma_1)  & -\lambda_c(1-\gamma_2)-\lambda \gamma_0 & \lambda_c \gamma_1               &    &   0 & 0\\
0                    & \lambda_c(1-\gamma_2)                   & -\lambda_c (1-\gamma_3+\gamma_1) &   &  \vdots & \vdots \\
\vdots               & 0                                      &  \lambda_c (1-\gamma_3)          &  \ddots  & \lambda_c \gamma_{n-1} &0   \\
0                     &                  \cdots                      &            0                      &   \lambda_c(1-\gamma_n) & -\lambda_c (1+\gamma_{n-1}) & 0 \\
0                     &                  \cdots                      &            0                      &  0 & G \lambda_c & - \nu 
\end{bmatrix}\label{eqn:jacobiKer}
$}
\end{split}
\end{equation} \end{linenomath}
where we omitted the argument $s$ for the sake of readability. We refer to $J_{f_1}(\gamma)$ as $J_{f_1}$ in the remainder.
\begin{theorem}\label{thm:JF1_neg}
For all $s\in(0,\bar{s})$ the Jacobian matrix of $f_1$ evaluated at $\gamma(s)$, has exactly one eigenvalue equal to zero and all remaining eigenvalues have real parts strictly smaller than zero. 
\end{theorem} 
\begin{prf}
First, we note that by applying the Laplace expansion for calculating the determinant of $(J_{f_1} -  \beta I)$ to obtain the characteristic polynomial, it becomes apparent that one eigenvalue $\beta$ is exactly equal to $-\nu$ and thus it remains to only study the eigenvalues of the matrix given by the first $n+1$ rows and columns of $J_{f_1}$ which we will call $J_{f_1}^{\text{red}}$.\\
Next, we decompose $J_{f_1}^{\text{red}}$ into the lower and upper triangular forms $J_{f_1}^{\text{red}} = L U$ with
\begin{linenomath} \begin{align}
L &=\begin{bmatrix}
1 & 0 & & \cdots & &  0\\
-\frac{1}{G}& 1 &  &  &  & \\
0& -1 & 1 & &  & \\
\vdots & & \ddots & \ddots  & & \vdots\\
&  & & -1 & 1 & 0\\
0 & \cdots & & 0 & -1 & 1
\end{bmatrix}\\
U &= \notag \\& \hspace{0cm}\resizebox{.44 \textwidth}{!}
{$
\begin{bmatrix}
-G \lambda(1-\gamma_1) & G \lambda \gamma_0                        & 0                                & \cdots   & 0 & G \lambda_c \\
0  & -\lambda_c(1-\gamma_2) & \lambda_c \gamma_1               &    &										 &   \lambda_c\\
\vdots        &       & -\lambda_c (1-\gamma_3) &  \ddots  & 														&  \vdots \\
               &          \ddots                            &       		 & \ddots  &   \lambda_c \gamma_{n-2} &  \lambda_c  \\
               &                                      &        &   &  -\lambda_c (1-\gamma_{n})  & \lambda_c (1+\gamma_{n-1})   \\
0                     &                  \cdots                      &                               &   &  0  &0
\end{bmatrix}
$}.
\end{align} \end{linenomath} 
As $\gamma_{i} \in (0,1)$ for $i=1,\ldots n$ by definition, we realize that all but the last diagonal entries of $U$ are strictly smaller than zero, whence the rank of $U$ is $n$. This proves that $J_{f_1}^{\text{red}}$ has exactly one zero eigenvalue.\\
We further find that $D=(L^{-1})^\top L^{-1}$ is positive definite and 
\begin{linenomath} \begin{align}
J_{f_1}^{\text{red}\top} D = U^{\top} L^{-1}.
\end{align} \end{linenomath} 
Now that $U^{\top}$ and $L^{-1}$ are lower triangular matrices and $L^{-1}$ has only ones on its diagonal, we conclude that $U^{\top} L^{-1}$ is also a lower triangular matrix and has the same diagonal entries as $U^{\top}$, therefore $U^{\top} L^{-1} \leq 0$. As this also holds for its transpose and it further holds that the sum of two negative semi-definite matrices remains negative semi-definite. Therefore $D$ solves the Lyapunov equation
\begin{linenomath} \begin{align}
 J_{f_1}^{\text{red}\top} D + D J_{f_1}^{\text{red}} = U^{\top} L^{-1} + (L^{-1})^\top U = Q \leq 0.
\end{align} \end{linenomath} 
Now, by applying Lyapunov's direct method \citep{Hahn1967}, we find that $\dot{\epsilon} = J_{f_1}^{\text{red}} \epsilon $ has a Lyapunov stable origin and therefore $J_{f_1}^{\text{red}}$ cannot have any eigenvalues with positive real part, completing the proof. \qed
\end{prf}
Similar to \cite{Halter2016}, we note that due to the zero eigenvalue of the Jacobian linearization the equilibria on the manifold $\gamma((0,\bar{s}))$ are non-hyperbolic. Therefore it is not possible to directly study the stability of the equilibria of the nonlinear system using its linearization as one would do by applying Lyapunov's indirect method \citep{Hahn1967} or more general the theorem of Hartman-Grobman \citep{Hartman1960}, that a vector field and its linearization are conjugate in a neighborhood of a hyperbolic equilibrium.\newline
We are thus left with studying non-hyperbolic fixpoints, e.g. by separately studying the restriction of our vector field to normal and tangent spaces of the submanifold of equilibria $\gamma((0,\bar{s}))$.\newline
As mentioned earlier, $f_1$ is normally hyperbolic at $\gamma((0,\bar{s}))$ if the Jacobian of $f_1$ evaluated at $\gamma(s)$ leaves the continuous splitting
\begin{linenomath} \begin{equation}
\mathbb{R}^{n+1} = N^u \oplus T_{\gamma(s)}\gamma((0,\bar{s})) \oplus N^s \label{eqn:splitting}
\end{equation} \end{linenomath}
invariant and if the normal behavior dominates the tangent one. Therein, $N^u$ and $N^s$ denotes the unstable and stable normal spaces of $\gamma((0,\bar{s}))$, i.e., the subspaces of the normal space spanned by the eigenvectors with positive and negative eigenvalues and $T_{\gamma(s)}\gamma((0,\bar{s}))$ for its tangent space.
\begin{lemma}\label{lem:lindep_eigenvector}
The eigenvector associated with the zero eigenvalue of $J_{f_1}$ is linearly dependent on $\dot{\gamma}(s)=\frac{d}{ds}\gamma(s)$.
\end{lemma}
\begin{prf}
It suffices to show that
\begin{linenomath} \begin{equation}
J_{f_1} \dot{\gamma} =0 \label{eqn: lem1_prf}
\end{equation} \end{linenomath}
holds for all $s\in(0,\bar{s})$. Therefore, we study each row of (\ref{eqn: lem1_prf}) separately, namely
\begin{linenomath} \begin{align}
J_{f_1}^0 \dot{\gamma} &=0 \\
J_{f_1}^i \dot{\gamma} &=0 \quad i = 1,\ldots, n-1\\
J_{f_1}^n \dot{\gamma} &=0 \\
J_{f_1}^{n+1} \dot{\gamma} &=0 
\end{align} \end{linenomath} 
with $J_{f_1}^i$ the $(i+1)$-th row of the Jacobian $J_{f_1}$. The latter two equations can be verified right away as the last three entries of $\dot{\gamma}$ are known explicitly. It remains to show the equality for an arbitrary row $i = 1,\ldots,n-1 $, which is given by
\begin{linenomath} \begin{equation}
\begin{split}
J_{f_1}^i \dot{\gamma} =&\lambda_c (1-\gamma_{i}) \dot{\gamma}_{i-1}\\
 - &\lambda_c (1-\gamma_{i+1}+\gamma_{i-1}) \dot{\gamma}_{i}\\
 + &\lambda_c \gamma_{i} \dot{\gamma}_{i+1}.
\end{split}
\end{equation} \end{linenomath} 
We rearrange the last equation to arrive at
\begin{linenomath} \begin{align}
\begin{split}
J_{f_1}^i \dot{\gamma} =& \lambda_c \left( \dot{\gamma}_{i-1} -  \gamma_{i}\dot{\gamma}_{i-1}-\gamma_{i-1}\dot{\gamma}_{i}  \right)\\
&- \lambda_c \left( \dot{\gamma}_{i} -  \gamma_{i+1}\dot{\gamma}_{i}-\gamma_{i}\dot{\gamma}_{i+1}  \right)
\end{split}\\
\begin{split}
=& \lambda_c \left( \dot{\gamma}_{i-1} - \dot{\overline{(\gamma_{i-1}\gamma_{i})}} \right)\\
&- \lambda_c \left( \dot{\gamma}_{i} - \dot{\overline{(\gamma_{i}\gamma_{i+1})}}\right).
\end{split}
\end{align} \end{linenomath} 
By studying the derivative of (\ref{eqn:gamma_recu}), we further realize that
\begin{linenomath} \begin{align}
\dot{\gamma}_{i} - \dot{\overline{(\gamma_{i} \gamma_{i+1})}} &=\begin{cases} \frac{\lambda_c}{\lambda} & i =0\\1 & i =1,\ldots,n-1 \end{cases} \label{eqn:gammadot}
\end{align} \end{linenomath} 
and using (\ref{eqn:gammadot}) for $i =1,\ldots,n-1$ we obtain 
\begin{linenomath} \begin{equation}
J_{f_1}^i \dot{\gamma} =0 \quad i =1,\ldots,n-1.
\end{equation} \end{linenomath}
Finally, we merely need to verify whether this is also true for the first row, i.e.
\begin{linenomath} \begin{align}
J_{f_1}^0 \dot{\gamma} &= -G \lambda(1-\gamma_{1})\dot{\gamma}_{0} + G \lambda \gamma_{0} \dot{\gamma}_{1} + G \lambda_c \dot{\gamma}_{n}\\
&= G \lambda (-\dot{\gamma}_{0} +\dot{\gamma}_{0} \gamma_{1} +\dot{\gamma}_{1} \gamma_{0} ) + G \lambda_c \dot{\gamma}_{n}.
\end{align} \end{linenomath} 
Now with $\dot{\gamma}_{n}=1$, and using equation (\ref{eqn:gammadot}) for $i=0$,
\begin{linenomath} \begin{align}
J_{f_1}^0 \dot{\gamma} &= G \lambda \left(-\frac{\lambda_c}{\lambda}\right)+ G \lambda_c =0.
\end{align} \end{linenomath} 
This concludes the proof. \qed
\end{prf}
We showed that the dynamics of $f_1$ on $T_{\gamma(s)}\gamma((0,\bar{s}))$ is determined by the zero eigenvalue and it remains to study the eigenvectors associated with the remaining eingevalues, showing that they span the normal space of $\gamma((0,\bar{s}))$ at any $\gamma(s)$.\newline
In \cite{Halter2016} we therefore introduced the affine subspaces $S_p$, which stemmed from the mass conservation of Ribosomes, showed that these subspaces are invariant under the flow of $f_1$ and further intersect transversely with $\gamma((0,\bar{s}))$. Accordingly we use the mass conservation law given by equation (\ref{eqn:CFM_fifth}) to find the $n$-dimensional subspaces $S_p(G)$. Due to the additional state $M$, we extend these subspaces by an additional basis vector to arrive at the $n+1$-dimensional subspaces
\begin{linenomath} \begin{equation}
S_p^{\text{ext}}(G)= \{e_1 p\} + \text{Im }\mu(G) \label{eqn:SpG}
\end{equation} \end{linenomath}
with $e_1$ the first vector of the standard basis of $\mathbb{R}^{n+2}$ and $\text{Im }\mu(G)$ the image of the matrix $\mu(G) \in \mathbb{R}^{(n+2) \times (n+1)}$ given by
\begin{linenomath} \begin{align}
\mu(G)= \begin{bmatrix}
-G & -G & \cdots & -G & 0\\
1 & 0 & \cdots & 0 & 0\\
0 & 1 &  &  \vdots & \vdots\\
\vdots & 0 &  & 0 & \\
0 & 0 & \cdots & 1 & 0\\ 
0 & 0 & \cdots & 0 & 1
\end{bmatrix}. 
\end{align} \end{linenomath} 
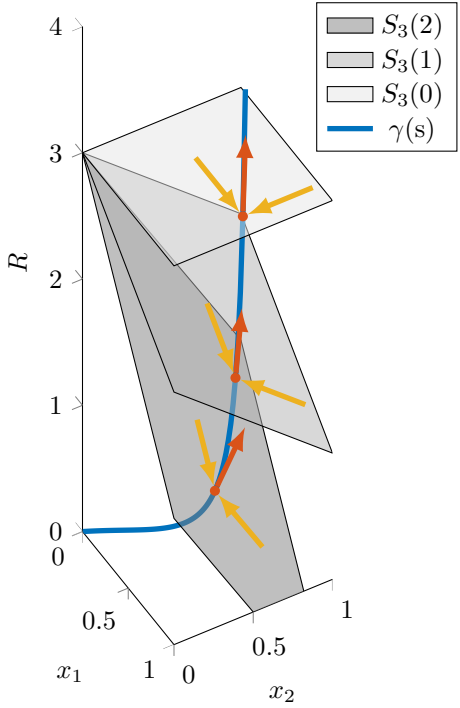
\begin{figure}
\centering
%
%

%
\begin{tikzpicture}

\begin{axis}[%
width=1.295in,
height=3.566in,
at={(2.371in,0.481in)},
scale only axis,
plot box ratio=1 1 4,
xmin=0,
xmax=1,
restrict x to domain=-1:2,
restrict y to domain=-1:2,
restrict z to domain=-1:6,
tick align=outside,
xlabel={$x_1$},
ymin=0,
ymax=1,
ylabel={$x_2$},
zmin=0,
zmax=4,
zlabel={$R$},
view={60}{46},
axis background/.style={transparent},
axis x line*=bottom,
axis y line*=left,
axis z line*=left,
legend style={at={(1.5,.94)}} 
]
\addplot3 [color=mycolor1,solid,line width=2.0pt,fill opacity=0.8,forget plot]
 table[row sep=crcr] {%
0	0	0\\
0.00507512175114543	0.00504949494949495	0.00253762623685843\\
0.0102020199958776	0.0100989898989899	0.00510154097250625\\
0.0153814911338734	0.0151484848484848	0.00769256555309408\\
0.0206143479820702	0.0201979797979798	0.0103115559005606\\
0.0259014201998953	0.0252474747474748	0.0129594043513829\\
0.031243554727783	0.0302969696969697	0.0156370416139303\\
0.0366416162394621	0.0353464646464646	0.0183454388534448\\
0.0420964876085254	0.0403959595959596	0.0210856099144622\\
0.0476090703898058	0.0454454545454545	0.0238586136913625\\
0.0531802853161137	0.0504949494949495	0.0266655566586973\\
0.0588110728109081	0.0555444444444444	0.0295075955739995\\
0.064502393517503	0.0605939393939394	0.0323859403669533\\
0.0702552288454324	0.0656434343434344	0.0353018572300856\\
0.0760705815346288	0.0706929292929293	0.0382566719275748\\
0.0819494762380945	0.0757424242424242	0.0412517733403461\\
0.087892960123778	0.0807919191919192	0.0442886172673786\\
0.0939021034963984	0.0858414141414141	0.0473687305050889\\
0.0999780004399912	0.0908909090909091	0.0504937152288181\\
0.106121769481989	0.095940404040404	0.0536652537028481\\
0.112334554279679	0.100989898989899	0.0568851133480519\\
0.118617524329931	0.106039393939394	0.0601551521992639\\
0.124971875703107	0.111088888888889	0.0634773247877899\\
0.131398831802135	0.116138383838384	0.0668536884882061\\
0.137899644147748	0.121187878787879	0.0702864103727665\\
0.144475593190948	0.126237373737374	0.0737777746214254\\
0.151127989153816	0.131286868686869	0.0773301905407375\\
0.157858172899803	0.136336363636364	0.0809462012508153\\
0.16466751683474	0.141385858585859	0.0846284931061918\\
0.171557425839823	0.146435353535354	0.0883799059239565\\
0.17852933823792	0.151484848484848	0.09220344410104\\
0.185584726794574	0.156534343434343	0.0961022887121492\\
0.19272509975519	0.161583838383838	0.10007981069078\\
0.199952001919923	0.166633333333333	0.104139585208146\\
0.207267015757883	0.171682828282828	0.108285407378983\\
0.214671762562344	0.176732323232323	0.112521309439324\\
0.222167903648727	0.181781818181818	0.116851579559739\\
0.229757141597219	0.186831313131313	0.121280782478618\\
0.237441221541977	0.191880808080808	0.125813782164315\\
0.245221932508971	0.196930303030303	0.130455766742748\\
0.253101108804618	0.201979797979798	0.135212275959158\\
0.261080631457465	0.207029292929293	0.140089231479778\\
0.269162429715323	0.212078787878788	0.145092970382036\\
0.27734848260032	0.217128282828283	0.150230282231729\\
0.285640820524541	0.222177777777778	0.155508450203523\\
0.294041526969002	0.227227272727273	0.16093529676872\\
0.302552740228881	0.232276767676768	0.166519234553301\\
0.311176655228085	0.237326262626263	0.172269323061958\\
0.319915525406374	0.242375757575758	0.178195332072965\\
0.328771664682457	0.247425252525253	0.18430781263741\\
0.337747449496656	0.252474747474747	0.190618176768707\\
0.346845320936915	0.257524242424242	0.197138787089183\\
0.356067786952162	0.262573737373737	0.203883057916091\\
0.365417424657232	0.267623232323232	0.210865569527083\\
0.374896882733795	0.272672727272727	0.218102197654381\\
0.384508883932005	0.277722222222222	0.225610260629287\\
0.394256227677817	0.282771717171717	0.233408687049049\\
0.404141792791221	0.287821212121212	0.241518207384835\\
0.414168540320945	0.292870707070707	0.249961573616373\\
0.424339516501477	0.297920202020202	0.25876381179547\\
0.434657855838623	0.302969696969697	0.267952513445746\\
0.445126784330155	0.308019191919192	0.277558172948887\\
0.455749622828507	0.313068686868687	0.287614579612904\\
0.466529790552882	0.318118181818182	0.298159275049187\\
0.477470808758566	0.323167676767677	0.309234088912707\\
0.488576304571734	0.328217171717172	0.320885769129569\\
0.4998500149985	0.333266666666667	0.333166726642676\\
0.51129579111755	0.338316161616162	0.346135919710831\\
0.522917602466204	0.343365656565657	0.359859909253236\\
0.53471954163043	0.348415151515152	0.374414125123655\\
0.546705829049948	0.353464646464646	0.389884394193538\\
0.558880818050265	0.358514141414141	0.406368795650099\\
0.571249000114273	0.363563636363636	0.423979928280675\\
0.583815010406784	0.368613131313131	0.442847700578315\\
0.596583633566316	0.373662626262626	0.463122789942698\\
0.609559809779295	0.378712121212121	0.48498096596824\\
0.622748641152906	0.383761616161616	0.508628540576259\\
0.63615539840384	0.388811111111111	0.534309303209975\\
0.649785527881375	0.393860606060606	0.562313435646939\\
0.663644658944443	0.398910101010101	0.592989098609573\\
0.677738611713666	0.403959595959596	0.626757673495578\\
0.692073405220816	0.409009090909091	0.664134079101536\\
0.706655265979637	0.414058585858586	0.705754250611234\\
0.721490637003702	0.419108080808081	0.752412910465872\\
};

\addplot3[area legend,solid,draw=black,fill=grey1,fill opacity=0.5]
table[row sep=crcr] {%
x	y	z\\
0	0	3\\
1	0	1\\
1	1	-1\\
0	1	1\\
}--cycle;
\addlegendentry{$S_3(2)$};

\addplot3 [color=mycolor1,solid,line width=2.0pt,fill opacity=0.8,forget plot]
 table[row sep=crcr] {%
0.721490637003702	0.419108080808081	0.752412910465872\\
0.736586187298714	0.424157575757576	0.805116427661625\\
0.751948821858658	0.429207070707071	0.865158299031551\\
0.76758569219727	0.434256565656566	0.934229415052097\\
0.783504207448562	0.439306060606061	1.0145833677153\\
0.799712046072628	0.444355555555556	1.10929176428825\\
0.816217168205559	0.449405050505051	1.22265242655447\\
0.833027828695217	0.454454545454545	1.36086912538559\\
0.850152590867722	0.45950404040404	1.53323985734852\\
0.86760034107287	0.464553535353535	1.75436077070718\\
};

\def \xa{0.7214}
\def \xb{0.4191}
\def \xc{0.7524}
\def \xaa{0.7066}
\def \xbb{0.4140}
\def \xcc{0.7057}
\def \G{2}

\node[circle, fill=mycolor3,scale=.4,outer sep=0pt] (A2) at (axis cs:0.7214,0.4191,0.7524){};
\node (B2) at (axis cs:0.9582,0.5007,1.4996){};
\draw[-latex, line width=0.7mm, mycolor3](A2)--(B2);

\node (C2) at (axis cs: 0.5014,0.4191,1.1924) {};
\draw[-latex, line width=0.7mm, mycolor2](C2)--(A2);

\node (D2) at (axis cs: 0.7214,0.7691,0.0524) {};
\draw[-latex, line width=0.7mm, mycolor2](D2)--(A2);

\addplot3[area legend,solid,draw=black,fill=grey2,fill opacity=0.5]
table[row sep=crcr] {%
x	y	z\\
0	0	3\\
1	0	2\\
1	1	1\\
0	1	2\\
}--cycle;
\addlegendentry{$S_3(1)$};

\addplot3 [color=mycolor1,solid,line width=2.0pt,fill opacity=0.8,forget plot]
 table[row sep=crcr] {%
0.86760034107287	0.464553535353535	1.75436077070718\\
0.885380304060424	0.46960303030303	2.04852676694671\\
0.903502059243137	0.474652525252525	2.45939199080146\\
0.921975557906794	0.47970202020202	3.07404966528938\\
};

\def \ya{0.8676}
\def \yb{0.4645}
\def \yc{1.7543}
\def \yaa{0.8501}
\def \ybb{0.4595}
\def \ycc{1.5332}
\def \yG{1}

\node[circle, fill=mycolor3,scale=.4,outer sep=0pt] (A) at (axis cs:0.8676,0.4645,1.7543){};
\node (B) at (axis cs:0.9201,0.4795,2.4176){};
\draw[-latex, line width=0.7mm, mycolor3](A)--(B);

\node (C) at (axis cs: 0.5176,0.4645,2.1043) {};
\draw[-latex, line width=0.7mm, mycolor2](C)--(A);

\node (D) at (axis cs: 0.8676,0.9645,1.2543) {};
\draw[-latex, line width=0.7mm, mycolor2](D)--(A);

\addplot3[area legend,solid,draw=black,fill=grey3,fill opacity=0.5]
table[row sep=crcr] {%
x	y	z\\
0	0	3\\
1	0	3\\
1	1	3\\
0	1	3\\
}--cycle;
\addlegendentry{$S_3(0)$};

\addplot3 [color=mycolor1,solid,line width=2.0pt,fill opacity=0.8]
 table[row sep=crcr] {%
0.921975557906794	0.47970202020202	3.07404966528938\\
0.940811141432201	0.484751515151515	4.09495576432047\\
0.960019560599251	0.48980101010101	6.1255080914873\\
0.979611996048774	0.494850505050505	12.1358252194364\\
0.999600079984003	0.4999	624.999975\\
};
\addlegendentry{$\gamma\text{(s)}$};

\def \za{0.9219}
\def \zb{0.4797}
\def \zc{3.074}
\def \zaa{0.9035}
\def \zbb{0.4746}
\def \zcc{2.4593}

\node[circle, fill=mycolor3,scale=.4,outer sep=0pt] (A3) at (axis cs:0.9219,0.4797,3.074){};
\node (B3) at (axis cs:0.9440,0.4858,3.8116){};
\draw[-latex, line width=0.7mm, mycolor3](A3)--(B3);

\node (C3) at (axis cs: 0.3219,0.4797,3.074) {};
\draw[-latex, line width=0.7mm, mycolor2](C3)--(A3);

\node (D3) at (axis cs: 0.9219,0.9797,3.074) {};
\draw[-latex, line width=0.7mm, mycolor2](D3)--(A3);


\end{axis}
\end{tikzpicture}%
 \caption{Three affine subspaces $S_3(0)$, $S_3(1)$ and $S_3(2)$ and the equilibria of (\ref{eqn:CFM_first})-(\ref{eqn:CFM_second}) given by $\gamma(s)$ depicted for the first three dimensions. Arrows visualize the linearized dynamics on the normal and tangent space of $\gamma((0,\bar{s})$.} \label{fig:SpG}
\end{figure}
Figure \ref{fig:SpG} depicts the first three components of such different $S_p^{\text{ext}}(G)$ for $p=3$ as well as $\gamma(s)$, the curve which represents all equilibria of (\ref{eqn:sysA}). In order to follow the same argumentation as in \cite{Halter2016}, we next show that any solution of (\ref{eqn:sysA}) initialized on $S_p^{\text{ext}}(G)$ will also remain on $S_p^{\text{ext}}(G)$ and subsequently state that $\gamma(s)$ intersects $S_p^{\text{ext}}(G)$ uniquely and transversely for all $p>0$ and $G \in \mathbb{N}$. With these statements at hand, we then realize that the continuous splitting given in equation (\ref{eqn:splitting}) exists, which is also depicted in Figure \ref{fig:SpG} by the linearized dynamics on the normal and tangent space of $\gamma((0,\bar{s}))$. 
\begin{lemma} \label{lem:Invariance_SpG}
All $S_p^{\text{ext}}(G)$ with  $p>0$ and $G \in \mathbb{N}$ are invariant sets of (\ref{eqn:sysA}).
\end{lemma}
\begin{prf}
For any values of $p$ and $G$ the vector 
\begin{linenomath} \begin{equation}
\mathbf{g}=\begin{bmatrix}
1&G&G& \hdots &G & 0
\end{bmatrix}^\top
\end{equation} \end{linenomath}
is perpendicular to $S_p^{\text{ext}}(G)$. Further, we note that
\begin{linenomath} \begin{align}
\langle f_1, \mathbf{g} \rangle = 0
\end{align} \end{linenomath}  which means that the vector field of the system given by (\ref{eqn:sysA}) always points in a perpendicular direction of $\mathbf{g}$. Therefore, the solutions of (\ref{eqn:sysA}) initialized in a certain $S_p^{\text{ext}}(G)$ cannot leave this subspace which concludes the proof. \qed
\end{prf}
\begin{lemma} \label{lem:Transversality_SpG}
For all $p>0$ and $G \in \mathbb{N}$ the curve $\gamma$ intersects $S_p^{\text{ext}}(G)$ transversely.
\end{lemma}
\begin{prf}
Using the same perpendicular vector $\mathbf{g}$ as above it suffices to show that the velocity vector of $\gamma$ is never perpendicular to $\mathbf{g}$ in order to conclude transversality of the intersection. As shown in \cite{Halter2016}, it holds that 
\begin{linenomath} \begin{equation}
\dot{\gamma}_i >0 \quad \forall i\in [0,n+1]
\end{equation} \end{linenomath}
and therefore
\begin{linenomath} \begin{equation}
\langle \dot{\gamma}, \mathbf{g} \rangle = \dot{\gamma}_0 + G \sum_{i=1}^n \dot{\gamma}_i >0,
\end{equation} \end{linenomath}
which concludes the proof. \qed
\end{prf}
With these lemmata at hand, we finally state our first result on the stability of the transcription model with multiple templates.
\begin{theorem} \label{thm:stab_sysA}
The invariant set $\gamma((0,\bar{s}))$ of (\ref{eqn:sysA}) is asymptotically stable.
\end{theorem}
\begin{prf}
With Lemmata \ref{lem:lindep_eigenvector}, \ref{lem:Invariance_SpG} and \ref{lem:Transversality_SpG} we conclude that $f_1$ is normally hyperbolic at $\gamma((0,\bar{s}))$ and therefore according to \cite{Pugh1970}, $f_1$ and the restriction of its linearization to the normal spaces of $\gamma((0,\bar{s}))$ are conjugate in a neighborhood of $\gamma((0,\bar{s}))$. In Lemma \ref{lem:lindep_eigenvector} we showed that the dynamics of $f_1$ restricted to the tangent space $T_{\gamma(s)}\gamma((0,\bar{s}))$ is determined by the zero eigenvalue of the Jacobian of $f_1$ while Theorem \ref{thm:JF1_neg} shows that the remaining eigenvalues are strictly smaller than zero. This reveals that the restriction of the linearization to the normal spaces of $\gamma((0,\bar{s}))$ has eigenvectors associated with eigenvalues with strictly negative real parts. Thus $\gamma((0,\bar{s}))$ is an asymptotically stable invariant set. \qed
\end{prf}

\subsection{Analysis of the complete model}
Above we found that Subsystem A of Figure \ref{fig:CFM_blocks}, governed by equation (\ref{eqn:sysA}), has a set of equilibria which is asymptotically stable. The output of this system is $M$, the amount of mRNA templates, simultaneously serving as input to Subsystem B of Figure \ref{fig:CFM_blocks}. In general, this means that while the template number of the just studied system is chosen to be static, the template number for Subsystem B varies with time. Now as $M$ is time dependent, the formerly used affine subspaces are not invariant under (\ref{eqn:sysA}) - (\ref{eqn:sysB}) anymore and it is thus not straight forward to study the convergence properties of Subsystem B independently of Subsystem A. For the overall protein synthesis model however, we are able to use the same argumentation as for the transcription part with the only difference that we now have to deal with a two dimensional manifold $\Psi$ representing the equilibria of the system. Let
\begin{linenomath} \begin{align}
\Gamma = \left\lbrace \begin{bmatrix}
\mathbf{x} \\ \mathbf{z} 
\end{bmatrix} \in \Omega_x \times \Omega_z \bigg\vert \begin{bmatrix}
f_1(\mathbf{x}) \\ f_2(\mathbf{x},\mathbf{z}) 
\end{bmatrix} =0 \right\rbrace
\end{align} \end{linenomath} 
be the set of equilibria of system (\ref{eqn:sysA}) - (\ref{eqn:sysB}). We define the two dimensional parameterization of these equilbria as
\begin{linenomath} \begin{align}
\begin{split}
\forall \begin{bmatrix}
\mathbf{x} \\ \mathbf{z} 
\end{bmatrix} \in \text{int }&(\Omega_x \times \Omega_z) \cap \Gamma \\
&\exists s \in (0,\bar{s}), l \in (0,\bar{l}) \quad \begin{bmatrix}
\mathbf{x} \\ \mathbf{z} 
\end{bmatrix} =\Psi(s,l) \label{eqn:Psidef}
\end{split}
\end{align} \end{linenomath} 
and find
\begin{linenomath} \begin{equation}
\Psi: (s,l) \mapsto \begin{bmatrix}
\gamma_{0}(s) \ldots & \gamma_{n+1}(s) & \xi_0(s,l) \ldots & \xi_{m+1}(s,l)
\end{bmatrix}^{\top}
\end{equation} \end{linenomath}
with the components $\gamma_{i}(s)$ given like in equation (\ref{eqn:gamma_recu}) and $\xi_{i}(s,l)$  similarly as
\begin{linenomath} \begin{equation}
\xi_{i}(s,l) = \begin{cases}
\tfrac{\eta_c l}{\eta(1-\xi_{1}(s,l))} & i = 0\\
 \tfrac{l}{1-\xi_{i+1}(s,l)} & i = 1 \ldots (m-1)\\
 l & i=m \\
\tfrac{G\lambda_c\eta_c}{\nu\delta}sl & i=m+1.
 \end{cases} \label{eqn:xi_recu}
\end{equation} \end{linenomath}
\begin{figure}
\centering
\begin{tikzpicture}
\node (A) at (0,0){};
\node (B) at (4,4){};
\node (C) at (2,.3){};
\node (D) at (6,4.3){};

\draw (A) to[out=20,in=200] node[anchor=east,midway]{$\gamma(\cdot)$} node[pos=.2,draw,inner sep=.05cm,fill=black,circle](S1){}  node[pos=.4,draw,inner sep=.05cm,fill=black,circle](S2){}  node[pos=.6,draw,inner sep=.05cm,fill=black,circle](S3){}  node[pos=.8,draw,inner sep=.05cm,fill=black,circle](S4){} (B);

\draw[opacity=0] (C) to[out=20,in=200] node[pos=.2](E1){}  node[pos=.4](E2){}  node[pos=.6](E3){}  node[pos=.8](E4){} (D);

\draw (S1) to[out=-10,in=200] node[anchor=north,midway]{$\xi(s_1,\cdot)$} (E1);
\draw (S2) to[out=-10,in=200] node[anchor=north,midway]{$\xi(s_2,\cdot)$} (E2);
\draw (S3) to[out=-10,in=200] node[anchor=north,midway]{$\xi(s_3,\cdot)$} (E3);
\draw (S4) to[out=-10,in=200] node[anchor=north,midway]{$\xi(s_4,\cdot)$} (E4);

\end{tikzpicture}%
\caption{Set of equilibria $\Psi$ as fiber bundle} \label{fig:fibers}
\end{figure}
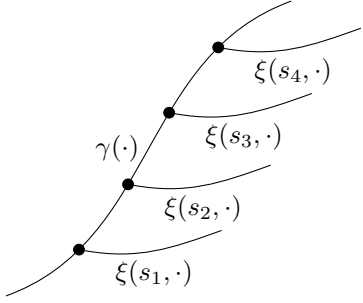
Geometrically, this is understood as follows: the manifold of equilibria, $\Gamma$, of (\ref{eqn:sysA}) - (\ref{eqn:sysB}) is two-dimensional, with $s$ and $l$ being local coordinates. But due to the cascaded structure depicted in Figure \ref{fig:CFM_blocks}, our equilibria $\Gamma$ are ``cascaded'' as well. In particular, $\Gamma$ cannot only be seen as a submanifold of $\mathbb{R}^{n+m+4}$, but also as a submanifold of the product state space $\mathbb{R}^{n+2} \times \mathbb{R}^{m+2}$. Taking this point of view we indeed find that $\Gamma$ is represented by the smooth fiber bundle  $\bigsqcup_{\gamma(s)} \xi(\lbrace(s,l)\vert l \in (0,\bar{l})\rbrace)$ as it is illustrated in Figure \ref{fig:fibers}. In other words, it is legitimate to think of $\Gamma$ as a one-dimensional submanifold, viz. the image of $\gamma$, with yet another one-dimensional submanifold $ \xi(\lbrace(s,l)\vert l \in (0,\bar{l})\rbrace)$ (the fibre, in the language of geometry), attached at every $\gamma(s)$. \newline
The Jacobian of the overall system evaluated at $\Psi(s,l)$ can now be found to be
\begin{linenomath} \begin{align}
J(\Psi) = \begin{bmatrix}
\tfrac{\partial f_1}{\partial \mathbf{x}} \big\vert_\Psi & 0 \\
&\\
\tfrac{\partial f_2}{\partial \mathbf{x}} \big\vert_\Psi& \tfrac{\partial f_2}{\partial \mathbf{z}}\big\vert_\Psi
\end{bmatrix}
\end{align} \end{linenomath} 
where $\frac{\partial f_1}{\partial \mathbf{x}}\big\vert_\Psi = J_{f_1}$ is known from equation (\ref{eqn:jacobiKer}), the off-diagonal block is given by
\begin{linenomath} \begin{equation}
\frac{\partial f_2}{\partial \mathbf{x}}\big\vert_\Psi = J_{f_2}^{\mathbf{x}}=  \resizebox{.27 \textwidth}{!}
{$
\begin{bmatrix}
0 		& \cdots & 	0	  & -G\lambda_c \sum_{i=1}^m \xi_i & \nu \sum_{i=1}^m \xi_i \\
 		& 		 & 		  & 0  							   & 0 \\
\vdots 	&  		 & \vdots & \vdots  					   & \vdots \\
 		& 		 & 		  & 0  							   & 0 \\
0 		& \cdots &  0     & 0  							   &  \eta_c \xi_m
\end{bmatrix}\label{eqn:jacobiOff}
$}
\end{equation} \end{linenomath}
and the lower diagonal block by
\begin{linenomath} \begin{equation}
\begin{split}
&\hspace{-0.4cm}\frac{\partial f_2}{\partial \mathbf{z}}\big\vert_\Psi = J_{f_2}^{\mathbf{z}}= \\ & \hspace{.2cm}\resizebox{.43 \textwidth}{!}
{$
\begin{bmatrix}
-\bar{M} \eta(1-\xi_1) & \bar{M} \eta \xi_0  & 0   & \cdots   & \bar{M} \eta_c & 0\\
\eta(1-\xi_1)  & -\eta_c(1-\xi_2)-\eta \xi_0 & \eta_c \xi_1               &    &   0 & 0\\
0                    & \eta_c(1-\xi_2)                   & -\eta_c (1-\xi_3+\xi_1) &   &  \vdots & \vdots \\
\vdots               & 0                                      &  \eta_c (1-\xi_3)          &  \ddots  & \eta_c \xi_{m-1} &0   \\
0                     &                  \cdots                      &            0                      &   \eta_c(1-\xi_m) & -\eta_c (1+\xi_{m-1}) & 0 \\
0                     &                  \cdots                      &            0                      &  0 & \bar{M} \eta_c & - \delta 
\end{bmatrix}\label{eqn:jacobiXi}
$}
\end{split}
\end{equation} \end{linenomath}
with $\bar{M}=\frac{G \lambda_c}{\nu}s$. Let $f:=\begin{bmatrix}
f_1 & f_2
\end{bmatrix}^\top$.
\begin{theorem}\label{thm:J_neg}
For all $s\in(0,\bar{s})$ and $l\in(0,\bar{l})$, $J(\Psi)$, the Jacobian matrix of $f$ evaluated at $\Psi(s,l)$, has exactly two eigenvalues equal to zero and all remaining eigenvalues have real parts strictly smaller than zero. 
\end{theorem}
\begin{prf}
Due to the block form of $J(\Psi)$, the eigenvalues of $J(\Psi)$ are given by the collection of the eigenvalues of $J_{f_1}$ and $J_{f_2}^{\mathbf{z}}$. In Theorem \ref{thm:JF1_neg} it was shown that $J_{f_1}$ has one eigenvalue equal to zero and all remaining ones have a real part strictly smaller than zero. We now note that $J_{f_2}^{\mathbf{z}}$ has exactly the same structure as $J_{f_1}$ and one can follow the same approach as in the proof of Theorem \ref{thm:JF1_neg} to show that the same statement holds for $J_{f_2}^{\mathbf{z}}$, therefore concluding the proof. \qed
\end{prf}
We now use the same approach as we took in \cite{Halter2016} and for the transcription model, namely that we first show that the overall protein synthesis model is normally hyperbolic at the manifold $\Gamma$ and subsequently restrict our attention to the dynamics on the normal spaces of $\Gamma$ in order to prove asymptotic stability of this manifold. \newline 
While for the transcription model, $ T_{\gamma(s)}\gamma((0,\bar{s}))$, the tangent space of the manifold of equilibria at a certain point $\gamma(s)$, was given by the span of $\frac{d}{ds}\gamma(s)$, the velocity vector of $\gamma(s)$, we are now facing a two dimensional tangent space given as
\begin{linenomath} \begin{align}
T_{\Psi(s,l)}\Gamma = \text{span} \left\lbrace \Psi_s(s,l) , \Psi_l(s,l) \right\rbrace
\end{align} \end{linenomath} 
where 
\begin{linenomath} \begin{align}
\begin{split}
\Psi_s & =\frac{\partial}{\partial s} \Psi(s,l) \\
&= \begin{bmatrix}
\dot{\gamma}_0 & \ldots & \dot{\gamma}_{n+1} & 0 & \ldots &0 & \frac{\partial}{\partial s} \xi_{m+1}
\end{bmatrix}^\top 
\end{split} \\
\begin{split}
\Psi_l&=\frac{\partial}{\partial l} \Psi(s,l) \\
&= \begin{bmatrix}
0 & \ldots &  0 & \frac{\partial}{\partial l}\xi_0 \ldots &\frac{\partial}{\partial l}\xi_{m+1}
\end{bmatrix}^\top.
\end{split}
\end{align} \end{linenomath} 

\begin{lemma}\label{lem:eig_PSI}
The two eigenvectors associated with the zero eigenvalues of $J(\Psi)$ are linearly dependent on $\Psi_s$ and $\Psi_l$ respectively.
\end{lemma}
\begin{prf}
In order to prove this Lemma, recalling Theorem \ref{thm:J_neg}, it suffices to show that $J(\Psi) \Psi_s = J(\Psi)\Psi_l =0$. Lets consider the first expression
\begin{linenomath} \begin{align}
J(\Psi) \Psi_s = \begin{bmatrix}
J_{f_1} \dot{\gamma} \\
J_{f_2}^{\mathbf{x}}\dot{\gamma} + J_{f_2}^{\mathbf{z}} \frac{\partial}{\partial s} \xi
\end{bmatrix}.
\end{align} \end{linenomath} 
As shown in the proof of Lemma \ref{lem:lindep_eigenvector}, we know that this first $n+2$ rows are all equal to zero and due to the structure of $J_{f_2}^{\mathbf{x}}$ and $\Psi_s$ it remains to study the first and last row of $J_{f_2}^{\mathbf{x}}\dot{\gamma} + J_{f_2}^{\mathbf{z}} \frac{\partial}{\partial s} \xi$. For the first row $J^{n+2}$, we find that
\begin{linenomath} \begin{align}
J^{n+2}(\Psi)\Psi_s &= -G\lambda_c \sum_{i=1}^m \xi_i \dot{\gamma}_n + \nu \sum_{i=1}^m \xi_i \dot{\gamma}_{n+1} \\
&= -G\lambda_c \sum_{i=1}^m \xi_i  + \frac{G \lambda_c}{\nu} \nu \sum_{i=1}^m \xi_i  = 0
\end{align} \end{linenomath} 
and similarly for the last row $J^{n+m+3}$ that
\begin{linenomath} \begin{align}
J^{n+m+3}(\Psi)\Psi_s &= \eta_c\xi_m \dot{\gamma}_{n+1} - \delta \frac{\partial}{\partial s} \xi_{m+1}\\
&= \frac{G \lambda_c \eta_c}{\nu} l - \delta \frac{G \lambda_c \eta_c}{\nu \delta} l = 0,
\end{align} \end{linenomath} 
revealing that $J(\Psi) \Psi_s =0$. Turning our attention to the second expression,
\begin{linenomath} \begin{align}
J(\Psi) \Psi_l = \begin{bmatrix}
0 \\
J_{f_2}^{\mathbf{z}} \dot{\xi}
\end{bmatrix}
\end{align} \end{linenomath} 
where we note that $J_{f_2}^{\mathbf{z}} \dot{\xi}$ is of the same structure as $J_{f_1} \dot{\gamma}$ and thus equality with zero can be shown in the same fashion as it was done in the proof of Lemma \ref{lem:lindep_eigenvector}. This reveals linear dependence of the two eigenvectors associated with zero eigenvalues on the vectors which span the tangent space of $\Gamma$ at $\Psi(s,l)$. \qed
\end{prf}
Now that we characterized the tangent space $T_{\Psi(s,l)}\Gamma$ it remains to show that the remaining eigenvectors span a space transversal to $\Gamma$ in order to conclude normal hyperbolicity of $\Gamma$. While for Theorem \ref{thm:stab_sysA} this was achieved by introducing the invariant affine subspaces $S_p^{\text{ext}}(G)$ explicitly, finding similar subspaces with the same properties for the whole model is not trivial. However, as we show briefly, an explicit characterization of the normal spaces is not needed to formulate our main result.
\begin{theorem}
The manifold of equilibria $\Gamma$ which is invariant of (\ref{eqn:sysA})-(\ref{eqn:sysB}) is asymptotically stable.
\end{theorem}
\begin{prf}
For the system being governed by equations (\ref{eqn:sysA})-(\ref{eqn:sysB}) to be normally hyperbolic at $\Gamma$, we need to find a continuous splitting of the state space into the tangent and normal spaces of $\Gamma$. With Lemma \ref{lem:eig_PSI} we showed that the eigenvectors associated with the zero eigenvalues span the tangent space of $\Gamma$ at any $\Psi(s,l)$. Due to Theorem \ref{thm:J_neg}, we also know that all other eigenvalues are different from zero and as eigenvectors belonging to pairwise distinct eigenvalues are always linearly independent \citep{Gantmacher1959}, we conclude that the remaining  eigenvectors span a subspace which is transversal to $\Gamma$ and those two subspaces together span the entire state space. Therefore, we again restrict our attention to the eigenvectors associated with the non-zero eigenvalues, which, as we showed in Theorem \ref{thm:J_neg}, have negative real parts, revealing that $\Gamma$ is asymptotically stable. \qed
\end{prf}

\subsection{Consequences for the biological system}
One may ask why the convergence properties of our model are of interest and some may even argue that a mathematical model as presented in Section \ref{sec:model} has its only purpose in generating predictions based on numerical simulations. While the property of asymptotic stability and the number of equilibria is indeed critical for evaluating the predictive power and the dynamic behavior of a mathematical model, the presented approach to study these properties additionally gives interesting insights from a system and control theoretic point of view. From this viewpoint, the major strength of mathematical models lies in their amenability to analytic methods with which one may assess different modes of manipulation, sensitivities of inputs and parameters or robustness of certain system outputs towards uncertain or disturbed parameters. In the present case where we found an asymptotically stable manifold of equilibria we infer that small disturbances in the system states and parameters do not affect the convergence towards this manifold of equilibria. However, the location of this manifold changes with variations in the parameters and the total amount of available resources and it is possible to use our model to analyze these changes. \newline
In particular, for the system at hand, we made the following system theoretic observations. Under the assumption of constant initiation of translation and transcription, the RNAP and Ribosome densities on the DNA and mRNA templates (locally) converge to certain fixed values which lie on a known and explicitly characterized two-dimensional submanifold, $\Gamma$. This convergence is exponentially stable, i.e. invariant under perturbations and the convergence rates can be approximated by the eigenvalues of the Jacobian. Therefore, given these rates, it is possible to characterize the different time scales of the protein synthesis model which may become crucial in the design process of synthetic gene networks. Specifically, $\Gamma$ can be efficiently computed as the nullspaces of $A(\mathbf{x})$ and $A(\mathbf{z})$, e.g. via the Gauss-Jordan-Algorithm. \newline 
Further interpreting the fiber bundle structure of $\Gamma$ as depicted in Figure \ref{fig:fibers}, we arrive at the following observations: If we hold the RNAP densities as well as amounts of mRNA and free RNAP fixed, then the Ribosome densities on the mRNA templates as well as amounts of protein and free Ribosomes may vary, namely along the fibers $\xi$, while leaving the overall system at rest. For a fixed initial amount of RNAP we can explicitly compute the eventual RNAP densities, amounts of mRNA and free RNAP, namely by intersecting the planes $S_p^{\text{ext}}$ with our curve $\gamma$. Thereafter, having that eventual values of RNAP densities, amounts of mRNA and free RNAP at hand, we can therefrom explicitly characterize the eventual Ribosome densities, amounts of protein and free Ribosomes in terms of the fibers $\xi$.

\section{Applications of the model}
Following the system theoretic analysis of the protein synthesis model we now focus on its predictive capabilities. In the remainder, we not only introduce a numerical example but further point out that this model can be used to capture the basal transcriptional and translational activity of an organism of interest. It may therefore be used to evaluate the (possibly limiting) effects of the finite pools of RNAP and Ribosomes on the performance of a synthetic genetic circuit, taking into account the housekeeping activity of a cellular system. We will refer to this approach as the background gene approach. In the following, we first introduce the relevant model parameters for simulating the basal protein production in E. Coli before we analyze the connection between transcription initiation and production rate and  ultimately show how the performance of a synthetic gene networks depends on the background gene activity. \newline
\subsection{Parameters for the basal activity in E. Coli} \label{sec:backgroundgene}
\begin{table*}[htbp]
\begin{center}
\begin{tabular}{llllrr}
Parameter & Property & Value & BioNum. ID & Publication\\
\hline
$L_G$ and $L_M$ & Avg. gene length & $1064$ $\si{nt}$ & 105751 & \cite{Rogozin2002}\\
$G$ & Expressed genes & $3000$ & 110942 & \cite{Tao1999}\\
$v_{\text{tx}}$ & RNAP speed of transcription & $3300$ $\si{nt\per\minute}$ & 111871 & \cite{Wang1998}\\ 
$R_{\text{rnap,tot}}$ & RNAP amount & $4600$ $\si{\per\cell}$ & 108601 & \cite{Bakshi2012} \\
$L_{\text{rnap}} $& RNAP size & $40$ $\si{nt}$ & 107873 & \cite{Selby1997}\\
$v_{\text{tl}}$ & Ribosome speed of translation & $2970$ $\si{nt\per\minute}$ & 100059 & \cite{Bremer2008} \\
$R_{\text{rib,tot}}$ & Ribosome amount & $39400$ $\si{\per\cell}$ & 101441 & \cite{Bremer2008}\\
$L_{\text{nt}}$ & Length of one nucleotide & $0.34$ $\si{\nano\metre\per nt}$  & 100667 & \cite{Langridge1960} \\
$L_{\text{rib,nm}} $ & Ribosome size & $26$ $\si{\nano\metre}$ & 100121 &\cite{Zhu1997} \\
$\nu$& mRNA degradation rate & $0.69$ $\si{\per\minute} $ & 111998 & \cite{Kennell1977} \\
$\delta $& Protein degradation rate & $\num{5.77e-4}$ $\si{\per\minute} $ & 111930 & \cite{Moran2013} \\
$\rho_{\text{rnap}}$ & RNAP engaged in transcription & $50$ \% & 110044 & \cite{Bakshi2013} \\
$\rho_{\text{rib}}$& Ribosomes engaged in translation & $80$ \% & 102344 & \cite{Bremer2008} \\
\\
\hline
Parameter & Property & Value & \multicolumn{2}{l}{Formula}  \\
\hline
$L_{\text{rib}} $ & Ribosome size& $76$ $\si{nt}$ &  \multicolumn{2}{l}{$L_{\text{rib,nm}} \cdot {L_{\text{nt}}}^{-1} $} \\
$n$ & DNA discretization points & $27$ & \multicolumn{2}{l}{$L_G \cdot {L_{\text{rnap}}}^{-1}$} \\
$m$ & mRNA discretization points  & $14$  &\multicolumn{2}{l}{$L_M \cdot {L_{\text{rib}}}^{-1}$}\\
$\lambda_c$ &Transcription elongation rate & $82.5$ $\si{\per\minute} $ &\multicolumn{2}{l}{$v_{\text{tx}} \cdot{L_{\text{rnap}}}^{-1}$}\\
$\eta_c$ &Translation elongation rate &$39.08$ $\si{\per\minute} $&\multicolumn{2}{l}{$v_{\text{tl}}\cdot{L_{\text{rib}}}^{-1}$}\\
$\lambda$ & Transcription initiation rate & $\num{1e-3}$ $\si{\per\minute} $ &\multicolumn{2}{l}{chosen}\\
$\eta$ &Translation initiation rate & $\num{1e-3}$ $\si{\per\minute} $ & \multicolumn{2}{l}{chosen}
\end{tabular}
\end{center}
\caption{Top: Average and typical values for parameters relevant for protein production in E. Coli obtained from the BioNumbers database \cite{Milo2010}. Bottom: Calculation of remaining parameters for E. Coli protein synthesis model.} \label{tab:paramDB}
\end{table*}

In order to simulate the basal protein production in E. Coli we use the model (\ref{eqn:CFM_first})-(\ref{eqn:CFM_sixth}), choose the parameters such that they represent an average gene of E.Coli and set the gene copy number of our model to the average amount of simultaneously expressed genes. As this resembles a reduction of all housekeeping genes to many copies of one single average gene, we termed this approach the background gene approach. \newline
The necessary parameters for this task were collected from the \emph{BioNumbers} database \citep{Milo2010} and converted into appropriate units (Table \ref{tab:paramDB}). We impose the simplifying assumption, that the gene length $L_G$ equals $L_M$, the length of the mRNA. Some of the parameters found in the \emph{BioNumbers} database need to be further processed for compliance with our protein synthesis model. These calculated parameters are collected in the bottom part of Table \ref{tab:paramDB}, including the formulae for how they are obtained. The initiation rates for both transcription and translation are the only parameters which were chosen freely in order to fit the model to average occupation values of RNAP ($\rho_{\text{rnap}}$) and Ribosomes ($\rho_{\text{rib}}$). \newline
\begin{figure}
\centering
\input{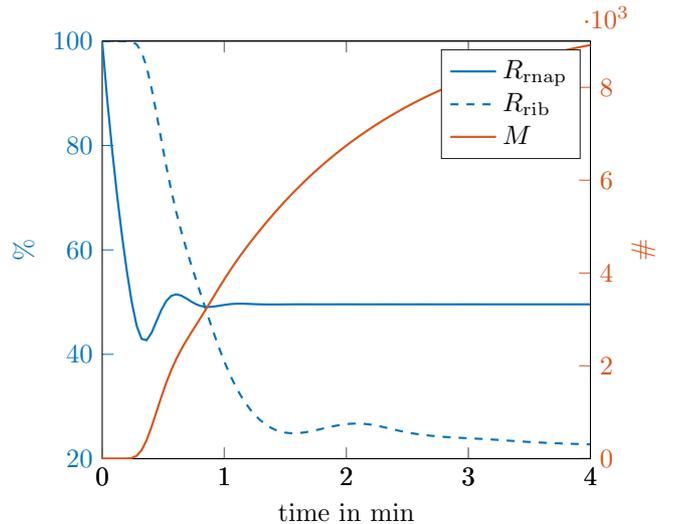} \caption{Amounts of unbound RNAP, $R_{\text{rnap}}$ (solid, blue, left scale), Ribosomes, $R_{\text{ribo}}$ (dashed, blue, left scale) and mRNA, $M$ (solid, red, right scale).} \label{fig:resources}
\end{figure}
Given these parameters, Figure \ref{fig:resources} depicts the simulated amounts of unbound RNAP and Ribosomes (in blue) as well as the amount of mRNA (in red). The system states of unbound RNAP and Ribosomes approach their steady-states (cf. plots of $R_{\text{rnap}}$ and $R_{\text{rib}}$ at $\sim 1$ and $\sim 3$ minutes respectively) and as pointed out before, the values of these steady states correspond to the values $\rho_{\text{rnap}}$ and $\rho_{\text{rib}}$ which were assessed from literature (see Table \ref{tab:paramDB}). This agreement was achieved by choosing the translation and transcription initiation rates appropriately. For the amounts of mRNA, we note that due to the fact that the RNAP has to travel through the DNA template first, a time delay becomes apparent at the beginning of the simulation. Further, the rate at which solutions approach their steady state is strongly dependent on the ratio of the initiation rate and mRNA degradation rate. \newline
With the steady state densities of RNAP and Ribosomes on the DNA and mRNA templates at hand, we also calculate the average amounts of RNAP and Ribosome units on a whole DNA and mRNA template respectively, viz.
\begin{linenomath} \begin{align}
\Phi_{\text{DNA}} &= \sum_{i=1}^n x_i(t\gg 0) = 0.7733 \\
\Phi_{\text{mRNA}} &= \sum_{i=1}^m z_i(t\gg 0) = 3.4166.
\end{align} \end{linenomath}
These values are again in accordance with statements from literature \citep{McAdams1997}, namely that in average there are usually several Ribosomes translating a single mRNA template at the same time, while only few RNAP are involved in the transcription of DNA.

\subsection{Nonlinear input-output behavior}
\begin{figure}[htbp]
\centering
\input{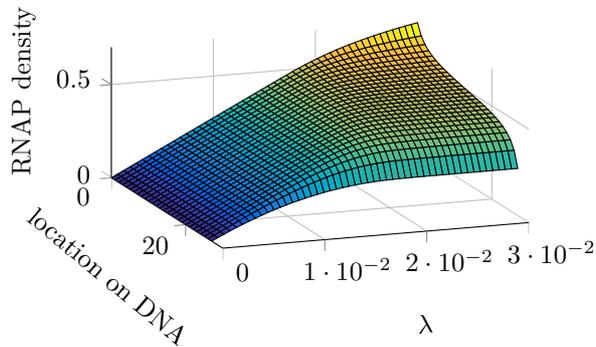} \caption{RNAP density on one DNA template for different transcription initiation rates $\lambda$.} \label{fig:nonlinIO}
\end{figure}
The main difference between our presented protein synthesis model and classic Hill-Kinetic or Mass-Action approaches is that besides the amounts of mRNA and protein, it is possible to observe the densities of RNAP and Ribosomes on the DNA and mRNA templates respectively. Therefore we obtain more accurate numbers on the density of the very last slot on a template. This slot in fact determines the real mRNA and protein production rates, thus our new model should yield more accurate results than classic approaches where this density is not considered at all. For the design of genetic interaction networks, it is important to characterize the input-output behavior of the parts of the network, where we consider e.g. transcription initiation rates as inputs and mRNA or protein amounts as output. We therefore study the relationships between the transcription initiation rate $\lambda$ of a gene (input) and the densities on the corresponding DNA template (output) and show the results of this numerical study in Figure \ref{fig:nonlinIO}. We find two interesting connections: While the density on the first slot (DNA location zero) for small $\lambda$ rises linearly with $\lambda$, it grows sub-linearly for larger values of $\lambda$ due to a limitation of the available amounts of RNAP. In contrast, the density on the last slot (DNA location $27$) shows a nonlinear behavior, very similar to a Michaelis-Menten kinetic, for the entire range of $\lambda$. This is a direct consequence of the transportation dynamics of RNAP and suggests that even in regimes where resources are available in abundance, one has to take the nonlinear input-output behavior between initiation and production rate for the design of networks into account.

\subsection{Load dependence of the repressilator}
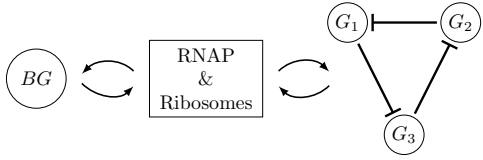
\begin{figure}
\centering
\resizebox{.35\textwidth}{!}{
\begin{tikzpicture}
\node[draw,circle,minimum size=30,inner sep=0pt,outer sep=8pt](BG) at (-5.5,1) {$BG$};
\node[draw,circle,minimum size=20,inner sep=0pt,outer sep=2pt](G1) at (0,2) {$G_1$};
\node[draw,circle,minimum size=20,inner sep=0pt,outer sep=2pt](G2) at (2,2) {$G_2$};
\node[draw,circle,minimum size=20,inner sep=0pt,outer sep=2pt](G3) at (1,0) {$G_3$};
\node[draw,rectangle,minimum width=2cm,inner sep=4pt,outer sep=8pt, align=center](pool) at (-2.5,1) {RNAP \\ \& \\ Ribosomes};
\node (hR) at (-.2,1) {};
\draw [-|,black,very thick](G1) -- (G3);
\draw [-|,black,very thick](G3) -- (G2);
\draw [-|,black,very thick](G2) -- (G1);
\draw [-latex,thick] (pool.174) to [out=135,in=45] (BG.6);
\draw [-latex,thick] (BG.354) to [out=315,in=225] (pool.186);
\draw [-latex,thick] (pool.6) to [out=45,in=135] (hR);
\draw [-latex,thick] (hR) to [out=225,in=315] (pool.354);
\end{tikzpicture}} \caption{Background gene and repressilator interacting with the same pool of RNAP and Ribosomes.} \label{fig:BG_approach}
\end{figure}
\begin{figure}
\centering
\input{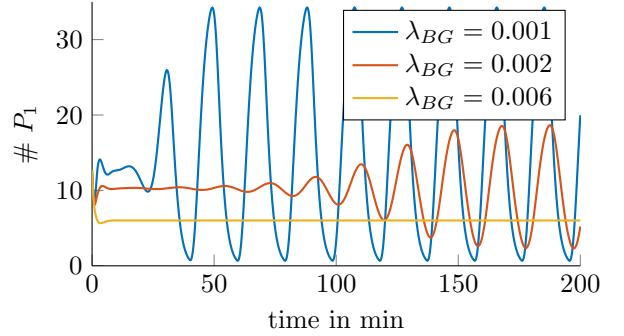} \caption{Trajectories of the amounts of one gene product of the repressilator, given three different environmental conditions.} \label{fig:trajectories}
\end{figure}

We motivated the presented model in order to more realistically simulate cellular environments and therefore evaluate the performance of a synthetic circuit with respect to changes in this environment. The repressilator, first introduced by \cite{Elowitz2000}, is a genetic circuit consisting of three genes repressing each other and quite popular in the field of synthetic biology. If the parameters of this circuit are chosen appropriately, one obtains oscillating behavior in the concentration of the gene products. The question how to choose these parameters to obtain this behavior in an isolated setting was mainly answered by \cite{Elowitz2000}. In \cite{Weisse2015} the same circuit was studied in terms of its performance with respect to the energy consumption in the host cell. With our presented model, we now can also study the effects of the limited pools of RNAP and Ribosomes on the performance of this specific circuit. In order to do so, we implemented the repressilator together with the background gene presented in Section \ref{sec:backgroundgene} as depicted in Figure \ref{fig:BG_approach}. The interactions between the genes $G_1$, $G_2$ and $G_3$ are modeled such that the transcription initiation rate of a gene $i$ depends on the protein amount of the $j$th gene via the Hill kinetic
\begin{linenomath} \begin{align}
\lambda_i = \lambda_i^{\text{basal}} - V \frac{P_j^h}{K^h+P_j^h}.
\end{align} \end{linenomath}
\begin{figure}
\centering
%
%
%
\begin{tikzpicture}

\begin{axis}[%
width=.35\textwidth,
height=2.7cm,
at={(0.758in,0.481in)},
scale only axis,
xmin=0.5,
xmax=3.5,
xtick={1,2,3},
xticklabels={{0.001},{0.002},{0.006}},
xlabel={$\lambda_{BG}$},
ymin=0,
ymax=60,
ylabel={\%},
axis background/.style={fill=white},
legend style={legend cell align=left,align=left,draw=white!15!black}
]
\addplot[ybar,bar width=0.229in,bar shift=-0.143in,draw=black,fill=mycolor1,area legend] plot table[row sep=crcr] {%
1	49.574971254965\\
2	32.962699866024\\
3	14.0872749352929\\
};
\addlegendentry{unbound RNAP};

\addplot[ybar,bar width=0.229in,bar shift=0.143in,draw=black,fill=mycolor3,area legend] plot table[row sep=crcr] {%
1	21.6079479906134\\
2	17.2263332775453\\
3	14.055896007114\\
};
\addlegendentry{unbound Ribosomes};

\end{axis}
\end{tikzpicture}%
 \caption{Unbound ressources RNAP and Ribosomes over basal transcription initiation rate $\lambda_{BG}$} \label{fig:diff_ressources}
\end{figure}
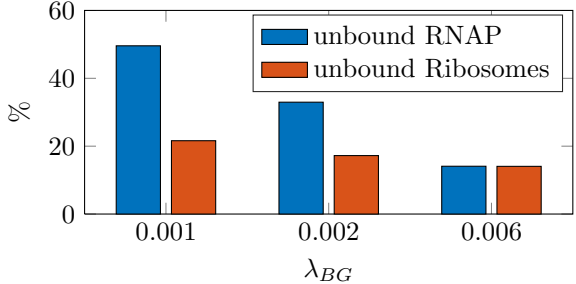
The choice for this kinetic is rather arbitrary and one can achieve similar results by using e.g. a linear function for the interaction between the genes. After finding some suitable parameters for the repressilator to ensure oscillating behavior, we modified the basal activity by changing the transcription initiation rate of the background gene. An increase in the basal transcription initiation rate $\lambda_{BG}$ results in more RNAP bound to the DNA, therefore producing more mRNA and thus also binding more Ribosomes, leaving less of the pool resources for the synthetic circuit. Figure \ref{fig:diff_ressources} depicts the unbound amounts of RNAP and Ribosomes for three different values of $\lambda_{BG}$. Therein, it becomes apparent how the amount of unbound RNAP and Ribosomes decreases with increasing basal transcription initiation and, in particular, that unbound RNAP decreases at a faster rate than unbound Ribosomes.\newline
For these different environmental conditions we implement the identical repressilator model and evaluate the amount of one gene product $P_{1}$. The simulation is initialized with the same initial conditions and Figure \ref{fig:trajectories} depicts the time evolution of $P_{1}$ for the different environmental conditions. While for the nominal system (blue) we obtained the desired oscillatory behavior, the amplitude  is not sustained when the transcriptional burden of the host cell is increased as is depicted for the red line. If it is increased even further (orange line), oscillations do no longer occur. This exemplifies the dependence of a synthetic circuit on the amount of available pool resources and the need for appropriate circuit designs which are robust with respect to such possible disturbances. The presented simulation framework may therefore be used as an in-silico test-bed or, using analytic and numeric methods, directly for the design of such robust circuits.


\section{Conclusion}
In order to better predict the performance of synthetic gene circuits we introduced a novel protein synthesis model based on the previously published RFM. In contrast to the typical approaches which are usually based on purely phenomenological observations, our model provides insight into the binding and movement of Ribosomes and RNAP on mRNA and DNA templates. Consequently, it is possible to evaluate the load of the transcriptional and translational processes on the limited pools of RNAP and Ribosomes in a cellular environment. Further, our model can be employed to also simulate networks of genes, specifically those interacting through both transcriptional and translational control mechanisms, and therefore providing a modeling framework for the design of mixed control networks.\newline
From a system theoretic point of view we showed that in order to analyze the convergence properties of the system, a geometric approach can be applied which is indifferent of the non-monotonic flow of the system. This proof further provides new insights into the dynamical properties and possible modes of manipulation. We found that the equilibria of the system can be characterized by a two-dimensional manifold which is asymptotically stable. Given a set of parameters and a fixed total amount of RNAP and Ribosomes, the system converges to a single equilibrium point. With the mathematical description of the manifold of equilibria at hand, we further know exactly how this steady state changes with respect to changing amounts of total RNAP and Ribosomes or different initiation rates of the gene. \newline
We showed that the presented model can be used to simulate the basal load of the transcriptional and translational machinery of e.g. E. Coli and that the obtained simulations are in accordance with data found in literature. We termed this approach the background gene approach and used this to study the input-output characteristics of a certain gene and further to evaluate the performance of the repressilator under different basal load scenarios. This revealed that the functionality of such a synthetic gene circuit is highly dependent on the basal load of the cell and we suggest to employ this framework to conduct a more detailed analysis of this resource dependence and eventually design more robust synthetic gene circuits.
\section*{Acknowledgments}
This work was supported by the research cluster $\text{BW}^2$ (www.bwbiosyn.de) of the Ministry for Science, Research and Art Baden-W\"urttemberg.


\bibliography{synthbio}

\end{document}